%% file: main.tex
\documentclass[conference]{IEEEtran}
\usepackage{subfiles}
\IEEEoverridecommandlockouts
\usepackage{cite}
\usepackage{amsmath,amssymb,amsfonts}
\usepackage{algorithmic}
\usepackage{graphicx}
\usepackage{textcomp}
\usepackage{xcolor}
\usepackage{url} 
\def\BibTeX{{\rm B\kern-.05em{\sc i\kern-.025em b}\kern-.08em
    T\kern-.1667em\lower.7ex\hbox{E}\kern-.125emX}}
\begin{document}

\title{A Survey of Information Disorder on Video-Sharing Platforms\\
}

\author{\IEEEauthorblockN{Meiyu Li}
\IEEEauthorblockA{\textit{University of Maryland} \\
College Park, USA \\
ml0521@umd.edu}
\and
\IEEEauthorblockN{Wei Ai}
\IEEEauthorblockA{\textit{University of Maryland} \\
College Park, USA \\
aiwei@umd.edu}
\and
\IEEEauthorblockN{Naeemul Hassan}
\IEEEauthorblockA{\textit{University of Maryland} \\
College Park, USA \\
nhassan@umd.edu}}

\maketitle

\begin{abstract}
Video-sharing platforms (VSPs) have become central information hubs but also facilitate the spread of information disorder, from misleading narratives to fabricated content. This survey synthesizes research on VSPs’ multimedia ecosystems across three dimensions: (1) types of information disorder, (2) methodological approaches, and (3) platform features. We conclude by identifying key challenges and open questions for future research.
\end{abstract}


\section{Introduction}
\subfile{sections/Introduction}

\section{Methodology}

\subfile{sections/Method}

\subfile{sections/Analysis_Results_ID_Types}

\subfile{sections/Analysis_Results_Methods}

\subfile{sections/Analysis_Results_Features}

\section{Challenges and Open Problems}
\subfile{sections/Challenges_and_Open_Problems}

\section{Related Work}
\subfile{sections/Related_Work}


\section{Conclusion}
\subfile{sections/Conclusion}

\bibliography{main.bib}{}
\bibliographystyle{IEEEtran}

\end{document}

%% file: sections/Introduction.tex
In recent years, video-sharing platforms (VSPs) have become central to how people consume entertainment, engage in public discourse, and access information. Platforms such as YouTube, TikTok, Instagram Reels, and X attract billions of daily active users, shaping how individuals encounter, process, and interpret content online \cite{west_167_2019}. While these platforms offer unprecedented opportunities for creative expression and global communication, they also serve as fertile ground for information disorder (ID)—the creation or dissemination of false or misleading information that may be either intentional or unintentional and potentially harmful \cite{wardle_information_nodate, zhao_prevalence_2023}.

The concept of information disorder, introduced by Claire Wardle, provides a more nuanced and comprehensive framework than the often-criticized term “fake news,” which has been dismissed for being overly vague and inadequate for capturing the diversity and complexity of false content online \cite{wardle_need_2018}. Unlike traditional text-based platforms, VSPs pose distinctive challenges for mitigating information disorder due to their rich multimedia format and algorithmically curated user experience. Prior studies show that the combination of visual, auditory, and textual elements in videos can enhance emotional appeal and persuasive power, making misleading content more compelling and memorable than text alone \cite{hornig2024misinformation, Exploring_the_Role, Chen_Kim_Gao_Raschka_2022}. In addition, algorithmic recommendation systems on VSPs often prioritize user engagement over information accuracy, fostering echo chambers that can amplify the spread of false or misleading content \cite{blackbox_audit, hornig2024misinformation}. Given the unique affordances of video-based platforms and the growing volume of research on this topic, there is a pressing need for a comprehensive survey that synthesizes current findings on ID in VSPs.

Although several literature reviews have addressed misleading information on VSPs, most are limited in scope. Many focus on a single subtype of ID, such as misinformation \cite{Combating_Online_Misinformation} or disinformation \cite{kapantai_systematic_2021}, or rely on vague and imprecise terms like ``fake news" that fail to capture the full complexity of the phenomenon \cite{sadman_covid-19_2024, a_survey_on_video, Multi-Modal_Misinformation_review}. In contrast, our survey organizes existing research using Wardle's typology of seven types of ID, providing a more granular and structured analysis that allows for nuanced examination of different forms of problematic content on VSPs \cite{wardle_understanding_2020}. Furthermore, while previous reviews tend to emphasize either computational approaches to misinformation detection \cite{Combating_Online_Misinformation, Multi-modal_Misinformation_Opportunity, Multimodal_Automated} or specific domains such as health misinformation on text-based social media \cite{zhao_prevalence_2023}, they overlook the broader ecosystem of platform features, audience behaviors, and recommendation mechanisms that contribute to the spread of false content. Our approach bridges these gaps by examining not only detection techniques but also the sociotechnical factors that shape information disorder in video environments, offering a holistic framework that connects content characteristics, user interactions, and algorithmic amplification patterns. In this paper, we synthesize literature spanning diverse domains, methodological approaches, and platform contexts, and we explore three key dimensions:
\begin{itemize}
\item \textbf{Types of Information Disorder}: The various forms of information disorder that appear on VSPs and how they have been conceptualized in existing research.
\item \textbf{Methodological Approaches}: The quantitative and qualitative approaches used to study information disorder in video environments.
\item \textbf{Features}: The features that contribute to the creation and spread of information disorder on VSPs.
\end{itemize}


%% file: sections/Method.tex
This research uses a systematic literature review with three components: structured keyword searches, multi-stage study screening, and a framework for categorizing ID types on VSPs.

\subsection{Paper Search and Keywords}
To ensure a comprehensive review, we conducted an advanced search on Google Scholar, known for its broad coverage \cite{gehanno2013coverage}. The search was limited to titles of publications from 2014 to 2024 and required at least one keyword from each of two curated sets: (1) the VSPs set included platform names such as \textit{YouTube} and \textit{TikTok}, along with general terms like \textit{video}, \textit{multimodal}, \textit{multimedia}, totaling 9 keywords. (2) The ID set comprised 43 keywords grounded in Wardle’s framework \cite{wardle_understanding_2020} (e.g., \textit{misinformation}, \textit{disinformation}, \textit{malinformation}, \textit{satire}, \textit{misleading}), and expanded based on prior surveys \cite{kapantai_systematic_2021, walter_how_2018, Combating_Online_Misinformation}, with terms like \textit{conspiracy}, \textit{rumor}, \textit{bias}, \textit{debunk}, and \textit{ambiguity}. We also included relevant linguistic variations (e.g., \textit{debunking}, \textit{ambiguous}) to maximize recall.

\subsection{Screening Process}
\begin{figure}[h]
  \centering
  \includegraphics[width=\linewidth]{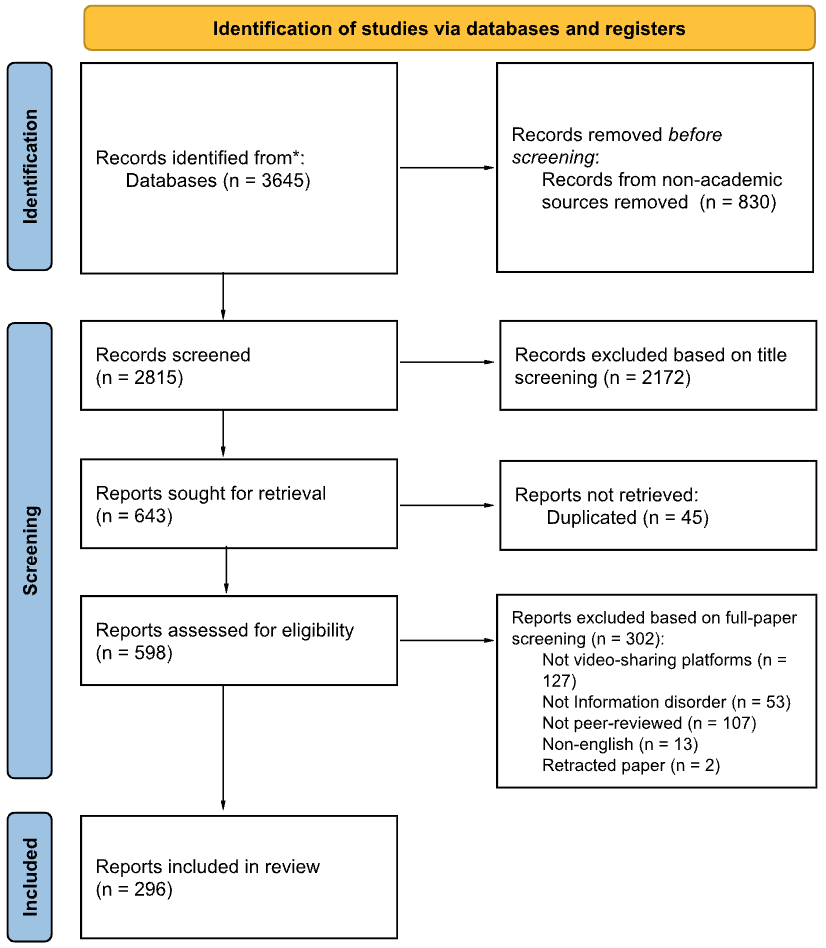}
  \caption{PRISMA diagram of literature identification, screening, and inclusion}
  \label{fig:prisma}
\end{figure}

The initial search using curated keyword sets yielded 3,645 papers. After removing non-academic sources (e.g., news articles), 2,815 academic papers remained. Fig. ~\ref{fig:prisma} outlines the four-step screening process. First, we excluded papers based on title review if they were unrelated to our topic despite containing relevant keywords (e.g., ``video games" or ``film video" instead of video-sharing platforms). Second, duplicates were removed. Third, full-text screening excluded papers that were: (1) not about VSPs \cite{10726230}, (2) unrelated to ID \cite{Nguyen02072024}, (3) not peer-reviewed \cite{2112.08611}, (4) non-English \cite{Busurkina_2021}, or (5) retracted \cite{2404.10702}. These steps yielded a final dataset of 296 papers.


\subsection{Information Disorder Categorization}


To classify the papers into established typologies of ID, we used a systematic content analysis approach grounded in Wardle's conceptual framework \cite{wardle_understanding_2020}. Our categorization was guided by a close reading of each paper’s research objectives, descriptions of misinformation, and the features of the datasets analyzed. Our coding relied on a conservative interpretive approach, assigning ID types when a clear and defensible alignment with Wardle’s framework could be inferred. If a paper exhibited characteristics spanning more than one category, we assigned multiple ID types to capture its multidimensional nature. Specific attention was paid to the source of misinformation, the modality of its presentation (e.g., title, audio), and the narrative or contextual framing. These elements were mapped onto the following seven types of IDs:

\begin{enumerate}
    \item \textit{Satire or Parody}: Videos that imitate news formats or documentary styles for humorous or critical purposes, without the intention to deceive, though viewers may misunderstand them as factual.

    \item \textit{False Connection}: Cases where video titles, thumbnails, or descriptions misrepresent the actual video content—for example, using sensational thumbnails (thumbnail bait) or headlines to attract clicks.

    \item \textit{Misleading Content}: Videos that present truthful elements in a deceptive way, such as through selective editing, emotionally charged narration, or framing that distorts the original context to mislead viewers.

    \item \textit{False Context}: Videos or clips that are repackaged with incorrect metadata, timestamps, or narrative framing, making old, unrelated, or geographically distant footage appear relevant to a current event.

    \item \textit{Imposter Content}: Videos that pretend to originate from credible sources—such as news outlets or public figures—by using falsified branding, voiceovers, or on-screen graphics to give a false sense of authority.

    \item \textit{Manipulated Content}: Videos containing genuine footage that has been digitally altered, such as splicing together unrelated clips, editing audio, or synthetic media techniques to misrepresent people or events.

    \item \textit{Fabricated Content}: Entirely false or staged videos created to deceive, often promoting hoaxes, conspiracy theories, or fictitious events that never occurred.
\end{enumerate}

%% file: sections/Analysis_Results_ID_Types.tex
\section{Types of Information Disorder}
Our analysis reveals variations in the prevalence of information disorders across VSPs. Notably, 289 out of 296 papers examine multiple types of information disorder simultaneously. For instance, studies focusing on clickbait videos typically address both misleading content and false connection \cite{Identification_of_Clickbait, identification_deceptive, kemm2022linguistic}, while research on conspiracy videos generally includes manipulated and fabricated content \cite{conspiracy_flat_earth, Antisemitic_conspiracy, Conspiracy_Theories_white}.

\begin{figure}[h]
  \centering
  \includegraphics[width=\linewidth]{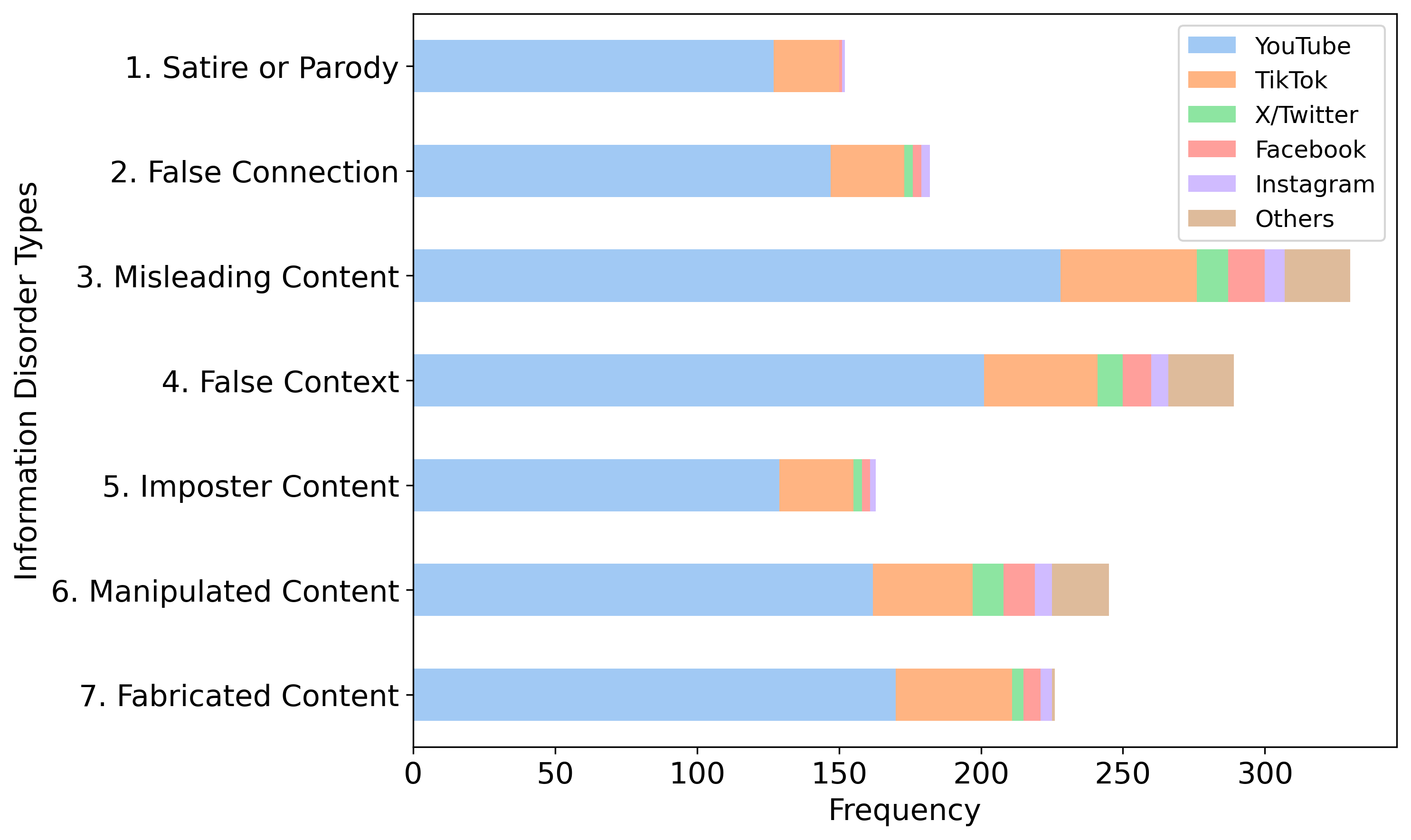}
  \caption{Distribution of information disorder types across VSPs}
  \label{fig:freq_ID}
\end{figure}

Fig.~\ref{fig:freq_ID}. illustrates the distribution of the seven types of information disorder, revealing a substantial imbalance across categories. The frequency counts reflect the aggregated occurrences of each type across the paper database, where a single paper may contribute to multiple categories. Misleading content and false context are the most frequently studied types, together accounting for approximately 39\%. In contrast, satire
and imposter content receives significantly less scholarly attention, comprising only 19.8\% \cite{Satire_and_Sarcasm, Satirizing_the_Clothing}.

This imbalance reflects a broader trend in misinformation research, where more overtly deceptive forms tend to attract greater focus, while subtler or more ambiguous categories are often overlooked \cite{wardle_need_2018}. For example, satirical content is frequently not perceived as misinformation because its original audience typically recognizes the humorous or parodic intent. However, when satire is reshared without attribution or contextual cues, it can easily be misinterpreted, contributing to the unintended spread of misinformation especially when encountered outside its original platform or by audiences unfamiliar with its context \cite{wardle_understanding_2020}. Overlooking these less-studied forms of information disorder may constrain our understanding of how users interpret and engage with content on VSPs, ultimately hindering the development of comprehensive strategies to address misinformation in digital environments. 

Additionally, Fig.~\ref{fig:freq_ID} reveals that research on short-form video platforms remains strikingly underdeveloped. While nearly half of all studies focus on YouTube’s long-form environment, only 27\% have examined short-form ecosystems. This gap is notable given that short-form video has become the dominant mode of online content circulation since the rise of TikTok (TT) in 2017 and the subsequent introduction of YouTube Shorts (2020), Instagram Reels (2020), and Facebook Reels (2021). Short-form platforms pose unique challenges for information disorder research \cite{fakesv, two_heads, Petrillo2021}. Their compressed narrative formats, rapid visual transitions, and fundamentally different patterns of user engagement distinguish them from long-form platforms: instead of deliberate information-seeking, users often consume content passively through endless scrolling \cite{Reshaping_digital_literacy, Environmental, sound_of_disinformation}. These behavioral differences require new approaches for detecting and analyzing misleading content. Moreover, researchers face significant technical and methodological barriers, including the difficulty of capturing context in brief clips, the role of algorithmically driven recommendation engines, and the challenges of analyzing fast-paced visual and audio dynamics \cite{Combating_Online_Misinformation, the_affective_algo, Reshaping_digital_literacy}. Addressing these obstacles is crucial for developing a more comprehensive understanding of how information disorder emerges and spreads within short-form video ecosystems.

A further challenge in understanding information disorder on VSPs is the difficulty of establishing standardized datasets that meet the diverse needs of researchers. As shown in TABLE~\ref{tab:dataset}, several publicly available datasets exist and are categorized according to the type of information disorder they address. However, only a few—such as FakeSV \cite{fakesv} focusing on Douyin, FVC \cite{fvc} (focused on YT(YouTube), FB(Facebook), X(Twitter/X)), VAVD (YT) \cite{misleading_metadata_detection}, MYVC (YT) \cite{CHOI202244}, and YTAudit1 (YT) \cite{measuring_misinformation}—have been reused in more than one study (typically two to four times). Most datasets are used only once, indicating limited generalizability and reusability.

This challenge stems from several factors. First, existing datasets are often narrow in scope—FakeSV \cite{fakesv}, for instance, focuses exclusively on Chinese (CHN) VSPs, while others like FVC \cite{fvc}, VAVD \cite{misleading_metadata_detection}, and MYVC \cite{CHOI202244} are either small in size or contain videos that have become inaccessible over time. Second, research questions vary widely—some studies investigate clickbait \cite{Detection_of_Clickbait}, conspiracy theories \cite{conspiracy_flat_earth}, or political misinformation \cite{Misleading_political_advertising}—yet most publicly available datasets cover only general categories of misinformation such as misleading content, false connection, or fabricated content \cite{FakingRecipe, Cross_Modal_Attention, Bias_Misperceived}, lacking the specific content or annotations required to support these focused inquiries. Third, some research explores platform-specific features, such as the behavior of recommendation algorithms \cite{blackbox_audit, hornig2024misinformation}, which require sequential or interactional data that most current datasets do not capture. These limitations underscore the need for both flexible, extensible datasets and domain-specific standardized resources tailored to particular research areas. Such datasets would enable greater consistency and comparability across studies while facilitating deeper analysis of specific types of information disorder and the dynamics of platform-specific content distribution.

\begin{table}[htbp]
\caption{Published dataset}
\begin{center}
\begin{tabular}{|l|l|l|r|}
\hline
\textbf{Dataset} & \textbf{ID Type} & \textbf{Platform} & \textbf{Size} \\
\hline
TripletViNet \cite{TripletViNet} & 2,3 & YT, X, Ins, FB & 100 videos \\
BaitBuster \cite{BaitBuster} & 2,3 & YT (Bangla) & 253,070 videos \\
YTDesign \cite{design_implications} & 2,3,4 & YT & 137,951 comments \\
YTConspiracy \cite{A_Longitudinal_Audit} & 3,4 & YT & 1,080 channels \\
TTUkraine \cite{fuzzy_approach} & 3,4 & TT & 16,000 videos \\
VMH \cite{sung-etal-2023-fake} & 3,4 & FB & 2,247 videos \\
YouNICon \cite{YouNICon} & 3,4 & YT & 596,967 videos \\
Fact-check \cite{Bias_Misperceived} & 3,4 & YT & 258 videos \\
YTReliable \cite{How_YouTube_Leads} & 3,4 & YT & 40,960 videos \\
$^{*}$FakeSV \cite{fakesv} & 3,4,6 & Others (CHN) & 3,654 videos \\
3M \cite{Cross_Modal_Attention} & 3,4,6 & Others (CHN) & 17,352 videos \\
$^{*}$FVC \cite{fvc} & 3,4,6 & YT, FB, X & 380 videos \\
$^{*}$VAVD \cite{misleading_metadata_detection} & 3,4,6 & YT & 650 videos \\
$^{*}$YTAudit1 \cite{measuring_misinformation} & 3,4,6 & YT & 56,475 videos \\
YTAudit2 \cite{An_Audit_of} & 3,4,6 & YT & 17,405 videos \\
FakeClips \cite{emotion_aided} & 3,4,6 & YT & 5,454 videos \\
FakeTT \cite{FakingRecipe} & 3,4,6 & TT & 1,991 videos \\
YTComments \cite{debating_in_echo_chambers} & 3,4,6,7 & YT & 1,199 comments \\
YTAuditConsp \cite{YouTube_and_Conspiracy} & 3,4,6,7 & YT & 620 search queries \\
$^{*}$MYVC \cite{CHOI202244} & 3,5,6 & YT & 287 videos \\
COVID-VTS \cite{liu-etal-2023-covid} & 3,4,6 & X & 10,000 videos\\
\hline
\multicolumn{4}{l}{$^{*}$ indicates this dataset has been used in more than one study.}\\
\multicolumn{4}{l}{ID Type column is defined in Section II.C.} \\
\end{tabular}
\label{tab:dataset}
\end{center}
\end{table}

%% file: sections/Analysis_Results_Methods.tex
\section{Methodological Approaches}
\begin{table}[htbp]
\caption{An overview of the methodological approaches.}
\begin{center}
\begin{tabular}{|l|r|}
\hline
\textbf{Method (n)} & \textbf{Description (n)} \\
\hline
Quant (80) & Machine learning method (25) \\
           & State-of-the-art models (21) \\
           & Statistical analysis (20) \\
           & Quant content analysis (7) \\
           & Data construction (3) \\
           & Social network analysis (3) \\
           & Mathematical modeling (1) \\
\hline
Qual (39)      & Qual content analysis (16) \\
               & User study (12) \\
               & Discourse analysis (9) \\
               & Ethnographic study (1) \\
               & Conceptual analysis (1) \\
\hline
Mixed (177) & Systematic medical video content analysis (142) \\
            & User study + statistical/content analysis (11) \\
            & Qual + quant content analysis (9) \\
            & Survey paper (8) \\
            & Social network analysis + Qual content analysis (6) \\
            & Audit study (1) \\
\hline
\end{tabular}
\label{tab:methods}
\end{center}
\end{table}


Having examined the 7 types of information disorder studied in VSPs, we now turn to how researchers have methodologically approached these phenomena. Table~\ref{tab:methods} provides an overview of the methodological approaches used in existing research. Quantitative approaches often examine a broad range of information disorder types without clearly specifying the particular misinformation topic under investigation. Qualitative and mixed-method approaches tend to focus on more narrowly defined types or topics, like conspiracy narratives in manipulated content, political satire, medical fabricated content.

\subsection{Quantitative} 
\begin{figure}[h]
  \centering
  \includegraphics[width=\linewidth]{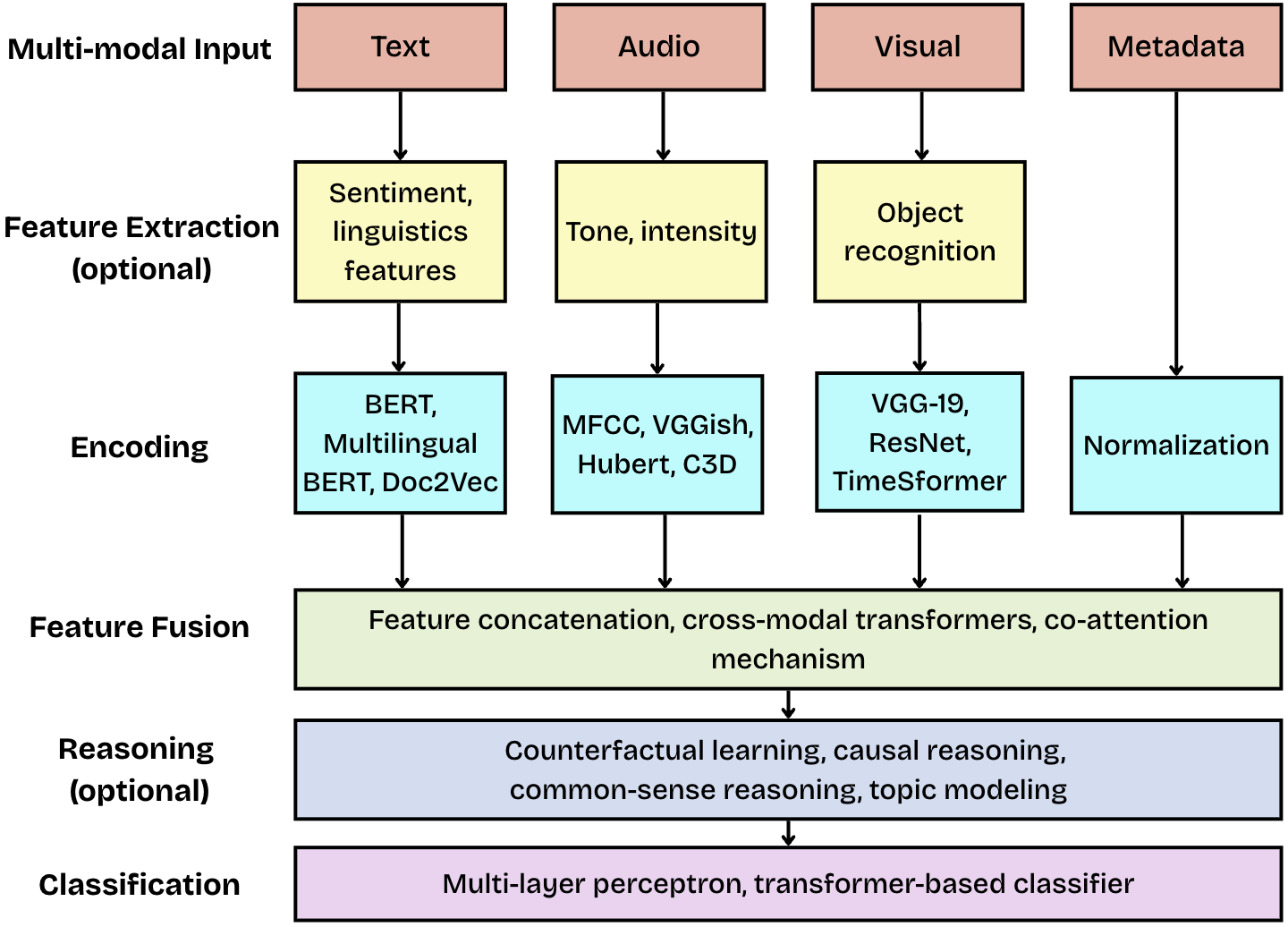}
  \caption{The architecture of SOTA models classifying misinformation videos}
  \label{fig:model}
\end{figure}

Quantitative methods are usually used to combat various ID types on VSPs by developing models that detect whether a video contains false content. For instance, 21 studies have proposed state-of-the-art (SOTA) models that process multi-modal features, such as visual, audio, and textual cues \cite{fakesv, FakingRecipe, wu-etal-2024-interpretable, MisD-MoE, mitigating_world_biases}. As illustrated in Fig.~\ref{fig:model}, most of these SOTA models share a common architecture consisting of three stages: multi-modal feature encoding, feature fusion, and final classification. Some studies enhance the feature encoding stage by integrating additional contextual signals, such as sentiment features derived from comments or audio tracks \cite{emotion_aided}. Others introduce an intermediate reasoning step like topic modeling to enrich contextual understanding before classification \cite{using_topic_modeling}.

Beyond classification tasks, other studies focus on analyzing the characteristics of misinformation videos or platform features. For example, most papers examine titles and thumbnails, key elements in false connection ID type, using quantitative content analysis like topic modeling or corpus analysis \cite{kemm2022linguistic}, video-language models \cite{sung-etal-2023-fake}, or machine learning approaches \cite{Detection_of_Clickbait, Identification_of_Clickbait, A_unified_approach}. A growing body of work also investigates the role of recommendation algorithms in promoting ID like misleading content or fabricated content. These studies often use audit methods to analyze the types of videos surfaced by YouTube's recommendation system and evaluate their alignment with misinformation narratives \cite{blackbox_audit, An_Audit_of, Assessing_enactment_of, measuring_misinformation}.

Despite their contributions, quantitative approaches face several limitations. First, SOTA detection models often rely on complex multi-modal architectures that can be difficult to interpret. This lack of interpretability limits transparency and hinders understanding of why a particular video is classified as misinformation. Furthermore, these models may struggle to generalize across different types of ID. For example, videos of ID types like satire or parody often include stylistic elements such as irony or humor that are difficult for models to detect and interpret \cite{10726230}. Similarly, manipulated content such as conspiracy theories may involve distinct narrative structures, tone, or audio features that differ significantly from the data used in existing SOTA models \cite{YouTube_and_Conspiracy}. Second, most quantitative studies focus on surface-level attributes, such as thumbnails \cite{Detection_of_Clickbait}, titles \cite{kemm2022linguistic}, and metadata \cite{misleading_metadata_detection}, while overlooking contextual and cultural factors that shape user perception and belief formation. This overemphasis on superficial features results in an incomplete understanding of how misinformation operates, spreads, and can be countered \cite{misinformation}. It underscores the need for more context-sensitive, interdisciplinary approaches that integrate sociocultural insights with quantitative methods.

\subsection{Qualitative and Mixed} In mixed methods, 142 medical papers use systematic content analysis to curate collections of medical videos from VSPs. These studies involve medical professionals who assess the informational quality of the videos using standardized instruments such as the DISCERN score \cite{charnock1999discern} and the JAMA benchmark criteria \cite{kloosterboer2019assessment}. The scope of these investigations spans a range of medical topics, such as COVID-19 \cite{li2022youtube, inaneroglu2022youtube}, vaccination \cite{abdaljaleel2024tiktok, gentile2023youtube}, and cancer \cite{breast_cancer_youtube, yalkin2022youtube, richartz2024canine}.

Another observation is the predominant use of qualitative and mixed-method approaches to examine specific topics within the information disorder. Many studies focus on fabricated content, especially conspiracy videos related to business or science \cite{Conspiracy_Theories_white, Antisemitic_conspiracy, picturing_opaque_power, Ha2022conspiracy, conspiracy_flat_earth}, using discourse or thematic analysis to unpack their narrative and rhetorical strategies. Other studies address domain-specific ID, such as political, climate, economic, scientific, or gender-related content \cite{America_First, from_denial_cultural, north_american, dangerous_but_efficacious, fake_news_dutch, Fake_news_as_fake_politics, sound_of_disinformation, Mind_over_matter, Environmental, Laughing_at_Trouble, Why_Are_Scientific_Experts, Anti-Women}, through similar methods, often integrating surveys or interviews to explore user perceptions and credibility judgments. This body of work underscores how misinformation is constructed and how it shapes user beliefs, attitudes, and trust on VSPs. Beyond topic-specific examinations, some propose interventions designed to counteract misleading and fabricated content \cite{seeing_is_not_believing, design_implications, from_adolescents, invid}, teaching media literacy \cite{media_and_information_literacy, teaching_critical_media} or evaluates the efficacy of existing fact-checking mechanisms \cite{Debunking_war_information_disorder}. These studies typically employ user-centered research methodologies like participatory design approaches.

These findings suggest that qualitative and mixed methods are well-suited for analyzing specific types of information disorder and for investigating individual misinformation events in greater depth. This methodological preference underscores the inherent complexity of information disorder phenomena, which cannot be fully captured through data-driven analysis or model development alone. However, such approaches are resource-intensive, often requiring substantial manual effort for data annotation, and they typically rely on small datasets, thereby limiting the generalizability of their findings.

%% file: sections/Analysis_Results_Features.tex
\section{Features}
To better capture the distinct research focuses, we categorized the features analyzed in existing studies into three major groups: content features, which relate to the characteristics of the media itself; user-engagement features, which reflect how users interact with the content; and algorithmic features, which involve the role of platform-driven mechanisms.

\subsection{Content features}

\begin{figure}[h]
  \centering
  \includegraphics[width=\linewidth]{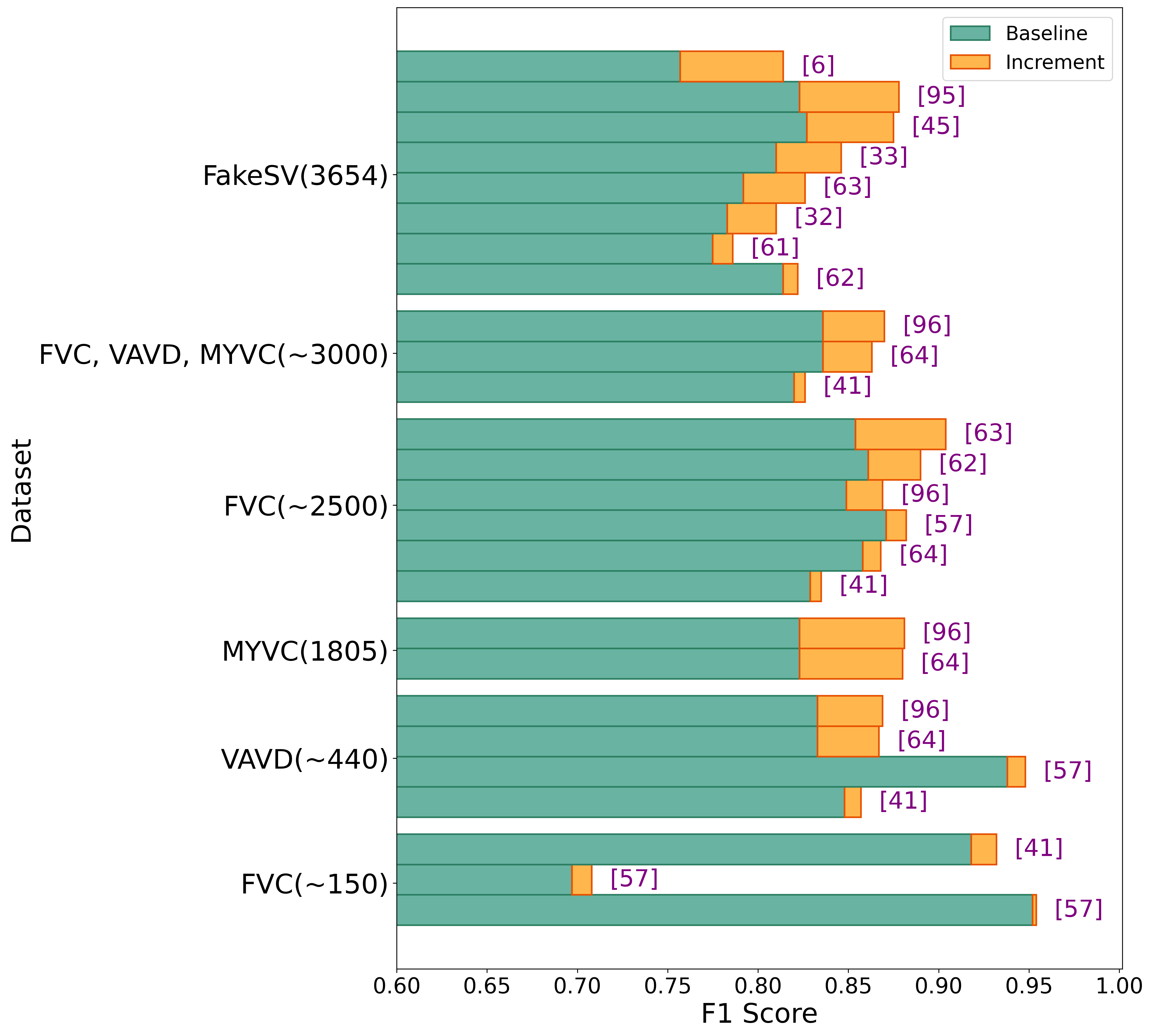}
  \caption{Comparison of F1 by SOTA multi-modal models}
  \label{fig:sota_f1}
\end{figure}

\begin{table}[htbp]
\caption{Prediction performance in current SOTA models}
\begin{center}
\begin{tabular}{|l|l|l|l|l|r|}
\hline
Paper&Text&Audio&Image&Video&Prediction score (F1)\\
\hline
\cite{fakesv} &Y&Y&Y&Y&0.81(0.027+)\\
\cite{two_heads} &Y&&&Y&0.85(0.036+)\\
\cite{unveiling_opinion} &Y&Y&Y&Y&0.88(0.055+)\\
\cite{FakingRecipe} &Y&Y&&Y&0.88(0.048+)\\
\cite{wu-etal-2024-interpretable} &Y&Y&&Y&0.79(0.011+)\\
\cite{mitigating_world_biases} &Y&Y&Y&Y&0.90(0.050+)\\
\cite{emotion_aided} &Y&Y&&Y&\textbf{0.95}(0.002+)\\
\cite{CHOI202244} &Y&&Y&Y&0.93(0.014+)\\
\cite{using_adversarial} &Y&&Y&Y&0.88(\textbf{0.058+})\\
\cite{using_topic_modeling}&Y&&Y&Y&0.88(0.057+)\\
\cite{Exploring_the_Role}&Y&Y&&Y&0.81(0.057+)\\
\cite{MisD-MoE}&Y&Y&Y&Y&0.89(0.037+)\\
\hline
\end{tabular}
\end{center}
\label{tab:sota}
\end{table}

We identified several content features that have been frequently examined in the literature including, titles, thumbnails, captions, audio, visual, specific content topics, narrative styles. Titles and thumbnails are commonly analyzed in false connection, often through the identification of linguistic markers such as personal and demonstrative pronouns (e.g., she, you, this), superlative adjectives, and intensifiers (e.g., most amazing), using corpus-based linguistic analysis \cite{kemm2022linguistic} or video-language models \cite{sung-etal-2023-fake}. In studying captions, audio, and visual features, most researchers use traditional machine learning methods and SOTA models \cite{fakesv, FakingRecipe, unveiling_opinion, two_heads, wu-etal-2024-interpretable, mitigating_world_biases, using_adversarial}. Fig.~\ref{fig:sota_f1} presents a comparative analysis of F1 scores achieved by multimodal SOTA models evaluated on standardized datasets. TABLE~\ref{tab:sota} lists the corresponding models shown in Fig.~\ref{fig:sota_f1}, detailing the highest F1 scores achieved and the specific modality combinations used by each model. The model proposed in \cite{emotion_aided} attained the highest F1 score (0.95) on the FVC dataset (150 videos) by integrating textual, auditory, and visual features. In contrast, the model in \cite{using_adversarial} demonstrated the most substantial improvement over the baseline on the MYVC dataset (440 videos) by jointly leveraging text, image, and video modalities. Furthermore, Table~\ref{tab:sota} reveals that nearly all SOTA models incorporate textual and visual features in their input, underscoring the significance of these modalities. In these models, textual features are typically extracted from transcripts and user comments, while visual features are from key frames in the videos.

Some studies also focus on content features like specific content topics or narrative styles, such as political discourse \cite{Assessing_enactment_of, Misleading_political_advertising, Fake_news_as_fake_politics, sound_of_disinformation}, environmental issues (e.g., climate change \cite{Environmental}), conspiracy-oriented narratives (e.g., flat earth theory \cite{conspiracy_flat_earth, Conspiracy_Theories_white, Antisemitic_conspiracy, YouTube_and_Conspiracy}), and satirical content in social events \cite{Satire_and_Sarcasm, Satirizing_the_Clothing, sixteen_not_pregnant}), and qualitative methodologies are commonly employed. These studies have identified rhetorical and stylistic characteristics of misinformation videos, including emotionally charged storytelling, the use of informal and relatable language, appeals to pseudo-authority or false expertise, and the presence of conspiratorial or anti-institutional frames.

\subsection{User-engagement features}
Among user-engagement features, the most frequently examined are comments and metadata (e.g., number of likes, views, and shares). In some studies, comments serve as the main contextual signal to support both traditional machine learning and SOTA models \cite{CHOI202244, fakesv, mitigating_world_biases, the_good_the_bad}. Other research conducts content analysis of user comments to examine discourse patterns and reactions to different types of information disorder \cite{analysis_user_generated_comments}, while additional studies analyze the structural properties of comment networks \cite{Misleading_political_advertising, Ha2022conspiracy} to better understand the dynamics of information disorder propagation. Beyond comments and metadata analysis, some studies assess users' judgment capabilities when encountering videos containing fabricated content, such as young audiences' evaluation skills \cite{Exploring_fake_news_awarness, from_adolescents, dangerous_but_efficacious} and users' digital literacy levels \cite{The_higher_the_news_literacy, Reshaping_digital_literacy, teaching_critical_media, media_and_information_literacy}. These investigations use participant recruitment for surveys and interviews, followed by qualitative coding of participants' experiences with VSPs.

\subsection{Algorithmic features}
In addition to the features above, some papers specifically evaluate recommendation algorithms and search engine built in video sharing platforms via quantitative methods. For example, several studies employ audit method to examine YouTube's algorithmic promotion of conspiracy content. These investigations reveal that filter bubbles form through recommendation mechanisms and watch history patterns, while personalization attributes exert limited influence \cite{YouTube_and_Conspiracy, the_affective_algo, dangerous_but_efficacious}. Other research applies machine learning to audit YouTube's recommendation systems. These studies develop specialized classifiers to detect videos containing fabricated content and systematically analyze recommendation trends. Findings suggest that advanced modeling techniques and collaborative filtering approaches demonstrate considerable potential for reducing the prevalence of information disorder within recommendation systems \cite{A_Longitudinal_Audit, hornig2024misinformation, Assessing_enactment_of}.

%% file: sections/Challenges_and_Open_Problems.tex
Our review identified key patterns in existing research and revealed three critical challenges that require further attention.
\subsection{Conceptual Limitation} A foundational limitation in current research on information disorder is the lack of nuanced definitions for its various forms. Many studies rely on the broad and often vague label of “fake news” to describe any type of misinformation, regardless of intent, format, or mechanism. However, “fake news” encompasses distinct subtypes of information disorder. For instance, it may include completely fabricated or manipulated videos designed to deceive, as well as satire or parody videos that were not originally intended to cause harm but nonetheless spread misinformation once circulated. Our analysis shows that attention to these different subtypes is uneven. For example, satire-related misinformation is studied far less frequently. To advance understanding of the “fake news” environment, different types of information disorder should be examined individually and addressed with targeted methods.

\subsection{Methodological Limitation} Another limitation in current research lies in the methodological approach. While SOTA models for detecting information disorder in videos have made progress by incorporating multimodal inputs, they continue to struggle with the complex, context-dependent nature of information disorder. These models rely on dataset-specific features and lack the adaptability needed to generalize across different platforms, topics, or cultural contexts. To better capture the nuances of information disorder on VSPs, it is essential to integrate qualitative insights (e.g., narrative framing, rhetorical strategies, or cultural cues) into quantitative modeling frameworks. This integration helps models move beyond surface-level signal detection toward a deeper understanding of meaning and intent. One promising direction for bridging this gap is the use of vision-language models (VLMs), which are trained on large-scale, diverse datasets that combine visual and textual content. VLMs provide a unified framework that jointly encodes multimodal information and supports generative reasoning, enabling more flexible and context-aware analysis \cite{VLM_multimodal}.

\subsection{Feature Coverage in Short Video Platforms} Although existing research provides broad coverage of features on VSPs, comparative analyses across platforms remain limited. For example, studies of recommendation systems and engagement features have predominantly focused on YouTube. However, with the rapid rise of short-video platforms, cross-platform comparisons are essential for identifying their unique features and understanding distinct underlying mechanisms. Unlike YouTube, short-video platforms such as TikTok do not use video titles, and users typically consume content by continuously scrolling. This unique mode of interaction warrants closer examination rather than relying solely on established feature categories derived from earlier platforms.

%% file: sections/Related_Work.tex
In 2017, Wardle introduced the concept of information disorder to replace the inadequate term fake news, offering a framework that distinguishes misinformation, disinformation, and malinformation, further detailed in a seven-part taxonomy from satire to fabricated content \cite{Giuliani-Hoffman2017, wardle_understanding_2020}. Several surveys have synthesized research on its characteristics, detection, and mitigation. Kapantai et al. reviewed disinformation typologies \cite{kapantai_systematic_2021}, while Walter and Murphy analyzed correction strategies \cite{walter_how_2018}. Domain-specific reviews in health, science, and politics \cite{wang_systematic_2019, chan_meta-analysis_2023, tucker_social_2018} often overlook the growing role of transmission media, particularly VSPs. These platforms promote multimodal content—combining visuals, audio, and text—through recommendation systems that privilege engagement over accuracy \cite{blackbox_audit}, encouraging passive consumption and virality that make misinformation harder to detect and easier to spread \cite{Exploring_the_Role, hornig2024misinformation}. While some surveys address deep learning–based detection on VSPs \cite{Multimodal_fake_news, Multi-Modal_Misinformation_review}, they often neglect algorithmic influence, user interaction patterns, and consistent taxonomies. Our work addresses this gap by adopting Wardle’s framework to analyze how different types of information disorder manifest on VSPs, offering a holistic and context-sensitive review.

%% file: sections/Conclusion.tex
This survey synthesizes research on ID across video environment via three key dimensions, identifies research gaps, and aims to inform future interdisciplinary efforts toward more effective detection and mitigation strategies.